\documentstyle[12pt]{article}

\catcode`\@=11
\long\def\@makefntext#1{
\protect\noindent \hbox to 3.2pt {\hskip-.9pt  
$^{{\ninerm\@thefnmark}}$\hfil}#1\hfill}		

\def\@makefnmark{\hbox to 0pt{$^{\@thefnmark}$\hss}}  
	
\def\ps@myheadings{\let\@mkboth\@gobbletwo
\def\@oddhead{\hbox{}
\rightmark\hfil\ninerm\thepage}   
\def\@oddfoot{}\def\@evenhead{\ninerm\thepage\hfil
\leftmark\hbox{}}\def\@evenfoot{}
\def\sectionmark##1{}\def\subsectionmark##1{}}

\renewcommand{\thefootnote}{\fnsymbol{footnote}}

\newcounter{sectionc}\newcounter{subsectionc}\newcounter{subsubsectionc}
\renewcommand{\section}[1] {\vspace*{0.6cm}\addtocounter{sectionc}{1} 
\setcounter{subsectionc}{0}\setcounter{subsubsectionc}{0}\noindent 
	{\normalsize\bf\thesectionc. #1}\par\vspace*{0.4cm}}
\renewcommand{\subsection}[1] {\vspace*{0.6cm}\addtocounter{subsectionc}{1} 
	\setcounter{subsubsectionc}{0}\noindent 
	{\normalsize\it\thesectionc.\thesubsectionc. #1}\par\vspace*{0.4cm}}
\renewcommand{\subsubsection}[1]
{\vspace*{0.6cm}\addtocounter{subsubsectionc}{1}
	\noindent {\normalsize\rm\thesectionc.\thesubsectionc.\thesubsubsectionc. 
	#1}\par\vspace*{0.4cm}}

\newcounter{appendixc}
\newcounter{subappendixc}[appendixc]
\newcounter{subsubappendixc}[subappendixc]

\renewcommand{\appendix}[1] {\vspace*{0.6cm}
        \refstepcounter{appendixc}
        \setcounter{figure}{0}
        \setcounter{table}{0}
        \setcounter{equation}{0}
        \renewcommand{\thefigure}{\Alph{appendixc}.\arabic{figure}}
        \renewcommand{\thetable}{\Alph{appendixc}.\arabic{table}}
        \renewcommand{\theappendixc}{\Alph{appendixc}}
        \renewcommand{\theequation}{\Alph{appendixc}.\arabic{equation}}
        \noindent{\bf Appendix \theappendixc #1}\par\vspace*{0.4cm}}

\def\abstracts#1{{
	\centering{\begin{minipage}{12.2truecm}\footnotesize\baselineskip=12pt\noindent
	\centerline{\footnotesize ABSTRACT}\vspace*{0.3cm}
	\parindent=0pt #1
	\end{minipage}}\par}} 

\renewenvironment{thebibliography}[1]
	{\begin{list}{\arabic{enumi}.}
	{\usecounter{enumi}\setlength{\parsep}{0pt}
\setlength{\leftmargin 1.25cm}{\rightmargin 0pt}
	 \setlength{\itemsep}{0pt} \settowidth
	{\labelwidth}{#1.}\sloppy}}{\end{list}}

\newcounter{itemlistc}
\newcounter{romanlistc}
\newcounter{alphlistc}
\newcounter{arabiclistc}

\newcommand{\fcaption}[1]{
        \refstepcounter{figure}
        \setbox\@tempboxa = \hbox{\footnotesize Fig.~\thefigure. #1}
        \ifdim \wd\@tempboxa > 6in
           {\begin{center}
        \parbox{6in}{\footnotesize\baselineskip=12pt Fig.~\thefigure. #1}
            \end{center}}
        \else
             {\begin{center}
             {\footnotesize Fig.~\thefigure. #1}
              \end{center}}
        \fi}

\newcommand{\tcaption}[1]{
        \refstepcounter{table}
        \setbox\@tempboxa = \hbox{\footnotesize Table~\thetable. #1}
        \ifdim \wd\@tempboxa > 6in
           {\begin{center}
        \parbox{6in}{\footnotesize\baselineskip=12pt Table~\thetable. #1}
            \end{center}}
        \else
             {\begin{center}
             {\footnotesize Table~\thetable. #1}
              \end{center}}
        \fi}

\def\@citex[#1]#2{\if@filesw\immediate\write\@auxout
	{\string\citation{#2}}\fi
\def\@citea{}\@cite{\@for\@citeb:=#2\do
	{\@citea\def\@citea{,}\@ifundefined
	{b@\@citeb}{{\bf ?}\@warning
	{Citation `\@citeb' on page \thepage \space undefined}}
	{\csname b@\@citeb\endcsname}}}{#1}}

\newif\if@cghi
\def\cite{\@cghitrue\@ifnextchar [{\@tempswatrue
	\@citex}{\@tempswafalse\@citex[]}}
\def\citelow{\@cghifalse\@ifnextchar [{\@tempswatrue
	\@citex}{\@tempswafalse\@citex[]}}
\def\@cite#1#2{{$\null^{#1}$\if@tempswa\typeout
	{IJCGA warning: optional citation argument 
	ignored: `#2'} \fi}}

 1
 1
 1

\font\ninerm=cmr9

\input psfig
\def\epem{e^+e^-}
\def\rts{\sqrt s}

\def\etal{{\it et al.}}
\def\eg{{\it e.g.}}

\def\fbi{~{\rm fb}^{-1}}

\def\wp{W^+}
\def\wm{W^-}

\newcommand{\alt}{\mathrel{\raisebox{-.6ex}{$\stackrel{\textstyle<}{\sim}$}}}

\def\lsim{\alt}

\def\gev{~{\rm GeV}}
\def\tev{~{\rm TeV}}

\def\ee{e^+e^-}

\def\mm{\mu^+\mu^-}
\def\mup{\mu^+}
\def\mum{\mu^-}

\begin{document}
\hspace*{\fill} FERMILAB-CONF-97/080-T\\
\hspace*{\fill} UCD--97--08 \\
\hspace*{\fill} April, 1997 \\
\bigskip
\centerline{\normalsize\bf STRONG WW SCATTERING PHYSICS:}
\baselineskip=20pt
\vskip -0.16in
\centerline{\normalsize\bf A COMPARATIVE STUDY FOR}
\baselineskip=20pt
\centerline{\normalsize\bf THE LHC, NLC AND A MUON COLLIDER
\footnote{Invited talk given at the Ringberg Workshop,
{\it The Higgs puzzle -- What can we learn from LEP II, 
LHC, NLC, and FMC?} Ringberg Castle, Germany, Dec. 8--13, 1996.}}
%

\centerline{\footnotesize TAO HAN}
\baselineskip=13pt
\centerline{\footnotesize\it Department of Physics,
University of California}
\baselineskip=12pt
\centerline{\footnotesize\it Davis, CA 95616}
\baselineskip=12pt
\centerline{\footnotesize\it and}
\centerline{\footnotesize\it Fermi National Accelerator Laboratory}
\baselineskip=12pt
\centerline{\footnotesize\it P.O.Box 500, Batavia, IL 60510}
\centerline{\footnotesize E-mail: than@ucdhep.ucdavis.edu}
\vspace*{0.3cm}

\vspace*{0.8cm}
\abstracts{We discuss the model independent parameterization 
for a strongly interacting electroweak sector. 
Phenomenological studies are made to probe such a sector
for future colliders such as the LHC, $e^+e^-$ 
Linear collider and a muon collider.}
 
\normalsize\baselineskip=15pt
\setcounter{footnote}{0}
\renewcommand{\thefootnote}{\alph{footnote}}
\section{Introduction}

The ``Higgs puzzle'' is clearly the most prominent question in
the contemporary high energy physics. 
Despite the extraordinary success of the Standard Model (SM) in
describing particle physics up to the highest energy available today,
the nature of electroweak symmetry-breaking (EWSB) remains undetermined.
Since any consistent EWSB theory will produce Goldstone bosons,
which become the longitudinal degrees of freedom of the electroweak
gauge bosons (generically denoted by $W_L$ unless specified 
otherwise), scattering of the Goldstone bosons would be the most
direct means to explore the nature of the EWSB. In particular, 
it is conceivable that there is no light ($\lsim 800\gev$)
Higgs boson. General arguments based on partial
wave unitarity then imply that the electroweak gauge
bosons develop strong (non-perturbative) interactions 
by energy scales of order 1--2~TeV and new physics beyond
the weakly coupled Higgs sector must show up\cite{sews}.
For a collider to probe such energy scales, 
the CM energy must be sufficient that gauge-boson scattering 
at subprocess energies at or above
1 TeV occurs with substantial frequency. 
The CERN Large Hadron Collider (LHC) would have the potential
to reach this physics. Other colliders under discussions 
that potentially meet this requirement are the
linear $\epem$ collider with $\rts \sim 1.5-2\tev$ (denoted
by NLC) and a high energy muon collider $\rts \sim 4\tev$.  

In this talk, we first discuss parameterizations
for a Strongly-interacting Electro-Weak Sector (SEWS)
in a (relatively) model-independent way. Based on this picture,
we then make a comparative study at
those future colliders for exploring the SEWS physics.

\section{Strongly Interacting Electroweak Sector}

Models with dynamical electroweak symmetry breaking 
often include strong self-interactions among the Goldstone bosons.
Technicolor theories are typical examples of this kind\cite{techni}.
Unfortunately, there are no phenomenologically successful models 
to describe such SEWS physics. To avoid conflict with the current
precision electroweak measurements, a phenomenological approach 
to study this class of physics is to only concentrate on the
minimal EWSB sector, and to parameterize the underlying physics
in a way independent of the detail dynamics by an effective Lagrangian, 
governed only by certain well motivated symmetries\cite{baggerL}. 

\subsection{``Model Independent'' Parameterization}

Motivated by the Standard Model electroweak gauge
symmetry $SU(2)_L \times U(1)_Y$ and realizing the approximate 
global symmetry $SU(2)_c$ (the custodial symmetry), one may
start from a global chiral symmetry $SU(2)_L \times SU(2)_R$.
Parameterizing the would-be Goldstone boson fields, 
$w^\pm$ and $z$ through the matrix 
\begin{equation}
\Sigma = \exp{(i \vec{\sigma}\cdot \vec{w}/v)}\ ,
\label{sigdef}
\end{equation}
we can construct an effective Lagrangian that contains this very general 
information:
\begin{equation}
{\cal L}= {v^2 \over 4} {\rm Tr}\partial_{\mu}\Sigma \partial^{\mu}
\Sigma^{\dagger}.
\label{lola}
\end{equation}
This is the lowest order in terms of the derivative expansion
and is model independent. Higher order terms
can be introduced systematically\cite{baggerL,hjhe}, 
but we choose to take this simplest case of Eq.~(\ref{lola})
as a representative.
The Goldstone boson scattering follows the Low energy 
theorem\cite{let} (LET) and there are no resonances
under consideration.
However, the scattering amplitudes in this case
violate the unitarity near 1--2 TeV region. 
For simplicity, we choose here to
unitarize them by the K-matrix technique, with the
model\cite{baggeretali} named LET-K.

Going beyond the non-resonance model, 
one can introduce a nonlinearly-realized 
chiral model with a spin-zero, isospin-zero resonance
($S$). The physics of this model 
is dominated by a new particle with the quantum numbers
of the Higgs boson.
For definitiveness in our presentation, we will choose
the scalar mass $M_S = 1~{\rm TeV}$ and width
$\Gamma_S = 350~{\rm GeV}$\cite{baggeretali}.
Similarly, one can study a nonlinearly-realized chiral model 
with a spin-one, isospin-one resonance\cite{bess,baggeretali}.  
The physics in this case is dominated by a 
(techni-$\rho$-like) vector particle.
We choose the mass-width combinations
$(M_V,\Gamma_V)=$ (1 TeV, 5.7 GeV) and (2.5 TeV, 520 GeV).
These models are fully described in Ref.~6.

\subsection{Low-lying States: Co-existence?}

The separate consideration on each individual low-lying 
state in a strongly-interacting sector may  be incomplete.
The well-known example
is the strong interactions of hadrons  in  which
the requirement of good high energy behavior
of  $S$-matrix elements gives rise
to a set of Adler-Weisberger (AW) type sum rules
for forward scattering amplitudes\cite{adlergilman}.
To satisfy  these sum rules,  besides the Goldstone bosons
$\pi's$, the other states such as $\rho, \sigma$, $a_1$
and $\omega$ are required to coexist.
It thus implies a scenario for the
strongly-interacting electroweak sector
which contains a rich spectrum of low-lying  states\cite{hhh},
including the Goldstone bosons ($w^\pm,z$ or  
equivalently the longitudinal components of the weak bosons
$W_L^{}$ in high energy limit\cite{sews}), a scalar resonance $H$,
a  $\rho$-like vector resonance $V$, and other vector
resonances such as $A_1$ and $\omega_H^{}$
(analogous to $a_1$ and $\omega$ in low energy hadron physics).
If the AW-type sum rules and the superconvergence relations
are applied to
$W_L^{}  W_L^{}$, $W_L^{} A_1$ and $W_L^{} V$ scatterings,
one obtains the relations among the couplings and masses, 
which are fully expressed
by two parameters: a mixing angle and one of the masses. This
scenario has significant phenomenological consequences\cite{hhh}.
A comprehensive study including all of the states at future
colliders is needed.

\section{Signatures of SEWS at Future Colliders}

In very high energy processes, there are many competing mechanisms
to produce $W$ bosons.  Some of these are more sensitive than
others to the EWSB sector involving the longitudinal components
$W_L$\cite{powerc}. 

The vector-boson scattering process,
$W_L^{}W_L^{} \to W_L^{}W_L^{}$, is clearly the most direct
means to explore the EWSB sector and is especially
important for the SEWS physics.
The major advantage for studying the vector-boson fusion processes
is that they involve all possible spin and isospin channels 
simultaneously, with scalar and vector resonances as well as 
non-resonant channels. For $s \gg M^2_W$, 
ignoring electroweak gauge couplings, 
the scattering of real longitudinal weak bosons 
is the same as the scattering of the corresponding
Goldstone bosons, in accordance to the Equivalence Theorem\cite{sews}.
In this case, the $WW$ scattering amplitudes can be parametrized 
by a single amplitude function\cite{baggeretali}. Furthermore,
there is a generic property of complementarity\cite{complmt,baggeretali}, 
namely: the
$W^+W^-$ and $ZZ$ channels should be more sensitive to a scalar
model; $W^\pm Z$ and $W^+W^-$ more sensitive to a vector model;
and $W^\pm W^\pm$ to a non-resonance model.

Another mechanism to produce vector boson pairs at future colliders is
through light fermion anti-fermion annihilation.  This is the case
of light $q \bar{q}$ annihilation in hadronic colliders
(the Drell-Yan type mechanism\cite{baggeretalii}), as well as
the case of $e^+ e^-$ annihilation in future $e^+ e^-$ 
colliders\cite{barklow}.
This production process is sensitive to new physics with a vector 
resonance intermediate state.

Finally, a mechanism for producing longitudinal vector boson pairs in
hadronic colliders is gluon fusion. In this case the
initial gluons turn into two vector bosons via an intermediate state
that couples to both gluons and electroweak gauge bosons. This process
may probe the SEWS physics in the heavy quark sector.
In our phenomenological discussions below, 
we will mainly concentrate on the first two mechanisms, 
due to the consideration of signal identification
and background suppression.

\subsection{The Large Hadron Collider}

The LHC should have good potential to probe the SEWS physics.
The major difficulty is the large SM backgrounds. In order to detect the
SEWS physics from the vector boson pair final state, we consider only
the clean channels, the so-called ``Gold plated'' pure leptonic
decay modes\cite{baggeretalii}. 
Table~\ref{lhcrates} presents the SM background
and signal rates for the models described above. We see that

\begin{itemize}

\item the isoscalar model gives rise to substantial
signal to background ratios 
in the $ZZ \to 4\ell$ and $ZZ \to 2\ell2\nu$ channels.
Especially encouraging is the signal rate for the  $2\ell 2\nu$ mode.  The
$W^+ W^-$ channel also exhibits some sensitivity to this model; the actual
sensitivity is probably somewhat greater since the distribution in the mass
variable $M(\ell\ell)$ peaks broadly around half the mass of the scalar resonance\cite{wpwm};

\item the isovector models (Vec~1.0 and Vec~2.5) gives substantial
signal in $q \bar q \to V \to W^\pm Z$ channel for $M_V \sim 1$ TeV,
and yield a continuum  event
excess in the $W^\pm W^\pm \to W^\pm W^\pm $ channel;

\item the non-resonant model (LET-K) yields observable
excesses in the $W^\pm W^\pm $ channel.

\end{itemize}

\begin{table}[tbh]
\centering
\caption{\label{lhcrates}
Event rates per LHC-year for $W_LW_L$ fusion signals from the different models,
together with backgrounds, assuming $\protect\sqrt s=14~{\rm TeV}$, 
an annual luminosity
of $100 ~{\rm fb}^{-1}$, and $m_t=175~{\rm GeV}$.  Cuts are listed in
Table~I of Ref.~12.
The $W^\pm Z(DY)$ row refers to the DY process
$q\bar q \to V \to W^\pm Z$,
with 0.85 $< M_T(WZ)<$ 1.05 TeV optimized 
for a 1 TeV vector state.}

\bigskip

\begin{tabular}{l|ccccc}
\hline\hline
& Bkgd. & Scalar & Vec~1.0 & Vec~2.5 & LET-K  \\
\hline
$Z Z(4\ell)$  & 0.7 & 4.6 & 1.4 & 1.3 & 1.4  \\
 \hline
$Z Z(2\ell2\nu)$ & 1.8 & 17 & 4.7 & 4.4 & 4.5  \\
 \hline
$W^+W^-$ & 12 & 18  & 6.2 & 5.5 & 4.6 \\
  \hline
$W^\pm Z$ & 4.9 & 1.5 & 4.5 & 3.3  & 3.0 \\
$W^\pm Z(DY)$ & 22 &  & 69 & & \\ \hline
$W^\pm W^\pm $ & 3.7 & 7.0 & 12 & 11 & 13 \\
\hline\hline
\end{tabular}
\end{table}

The results clearly demonstrate that one should examine all vector boson
pair channels and search for the deviation from the SM.
From Table~\ref{lhcrates} for the $W_LW_L$ signals
and the predicted background rates,
we can estimate the number of LHC years
necessary to generate a signal 
at the 99\% Confidence Level\cite{baggeretalii}.
This is given in Table~\ref{lhclum}.
We see that with a few years running at the LHC, 
one should be able to
observe a significant enhancement in at least one gold-plated channel.
Such an enhancement would be an important step towards revealing the
physics of strongly interacting electroweak symmetry breaking.

More detailed simulations\cite{atlas} 
agreed with our conclusions presented here. 
In particular, considering the semi-leptonic decay modes
for a 1 TeV Higgs boson $H\to WW \to \ell \nu jj$, the signal 
observability can be substantially increased\cite{atlas,cms}.

\begin{table}[tbh]
\centering
\caption{\label{lhclum}
Number of years (if $<10$) at LHC required for a
99\% confidence level signal.}
\bigskip

\begin{tabular}{l|cccc}
\hline\hline
& \multicolumn{4}{c}{Model}\\ \cline{2-5}
Channel& Scalar & Vec~1.0 & Vec~2.5 & LET K \\
\hline
$ZZ(4\ell)$     & 2.5  & \    & \     &       \\
$ZZ(2\ell2\nu)$ & 0.75 & 3.7  & 4.2   & 4.0   \\
$W^+W^-$        & 1.5  & 8.5  & \     & \     \\
$W^\pm Z$       & \    & 0.07 &       &       \\
$W^\pm W^\pm $  & 3.0  & 1.5  & 1.5   & 1.2   \\
\hline\hline
\end{tabular}
\end{table}

\subsection{TeV $e^+e^-$ Linear Collider}

Due to the cleaner experimental environment at the NLC, 
it may be desirable to
study the hadronic decay modes to enlarge the signal sample for
the $WW$ final state.
Our approach is based on $W^+W^-,ZZ\to (jj)(jj)$ four-jet signals, and
therefore relies on good dijet mass resolution\cite{sewsnlc}. 

Our main results are summarized in Table~\ref{tableii}
for an $e^+e^-$ (and $e^-e^-$) collider at $\sqrt s=1.5$~TeV.
They show that 
\begin{itemize}

\item a 1-TeV scalar or vector state would be easily
observable through the process $W^+W^- \to ZZ$ and $W^+W^-$,
respectively;

\item a 5.7$\sigma$ signal for the LET amplitudes can be obtained
via the $W^+W^- \to ZZ$ channel alone;

\item 
the $W^+W^-/ZZ$ event ratio is a sensitive probe of SEWS dynamics.
Indeed, the differences between the various models are quite
marked: a broad Higgs-like scalar will enhance both $W^+ W^-$
and $ZZ$ channels with $\sigma(W^+ W^-) > \sigma(ZZ)$; a $\rho$-like 
vector resonance will manifest itself through $W^+W^-$ but 
not $ZZ$; the LET-K amplitude will enhance $ZZ$ more than $W^+ W^-$;

\item 
for an $e^-e^-$ collider with the same energy and luminosity
(see the last row in Table~\ref{tableii}),
the LET signal rate  for the $\nu\nu W^-W^-$ ($I=2$) channel
is similar to the LET result of  $e^+e^-\to\bar\nu\nu ZZ$, as anticipated,
while the background rate is higher.

\end{itemize}

For a given luminosity, 
the signals are doubled for an $e^-_L$ polarized beam (or
quadrupled for two $e^-_L$ beams), whereas the backgrounds increase
by smaller factors.  Hence polarization improves the significance
of signals substantially\cite{sewsnlc}.

The direct $s$-channel process $e^+e^- \rightarrow W^+W^-$
should be more advantageous in searching for effects from a vector
$V$ through $\gamma,Z-V$ mixing, due to
more efficient use of the CM energy, the known beam energy constraint,
and better control of backgrounds. For instance, at a 1.5 TeV NLC
with 225 fb$^{-1}$ luminosity, a 7$\sigma$ effect may be obtained
for a 4 TeV vector state\cite{barklow,nlc}. 
Nevertheless, the $WW$ fusion
processes studied here involve more  spin-isospin channels
of $WW$ scattering; they are unique
for exploring scalar resonances and are complementary to the direct
$s$-channel for the vector and non-resonant cases.

\begin{table}[tbh]
\centering
\caption[]{\label{tableii}\small
Total numbers of $W^+W^-, ZZ \rightarrow  4$-jet
signal $S$ and background $B$ events calculated for  a 1.5~TeV
NLC with  integrated luminosity 200~fb$^{-1}$.  Events are summed
over the mass range $0.5 < M_{WW} < 1.5$~TeV except for the $W^+W^-$ channel
with  a narrow vector resonance in which $0.9 < M_{WW} < 1.1$~TeV. The
statistical significance $S/\sqrt B$ is also given.
For comparison, results for $e^-e^- \rightarrow \nu \nu W^-W^-$
are also presented, for the same energy and luminosity and the $W^+W^-$
cuts. The hadronic branching fractions of $WW$ decays and the $W^\pm/Z$
identification/misidentification are included (see Ref.~17 for more details).}
\medskip
\begin{tabular}{|l|c|c|c|c|} \hline
channels &  Scalar & Vector   & LET  \\
\noalign{\vskip-1ex}
& $M_S=1$ TeV & $M_V=1$ TeV &\\\hline
$S(e^+ e^- \to \bar \nu \nu W^+ W^-)$
&  160   & 46  & 31  \\
$B$(backgrounds)
& 170   & 4.5  & 170  \\
$S/\sqrt B$ & 12 & 22 & 2.4 \\
\hline
$S(e^+ e^- \to \bar\nu \nu ZZ)$
&  130  & 36  & 45   \\
$B$(backgrounds)
& 63     & 63  & 63  \\
$S/\sqrt B$ & 17& 4.5& 5.7\\
\hline
\hline
$S(e^- e^- \to \nu \nu W^- W^-)$
&  35  & 36  & 42  \\
$B$(backgrounds)
& 230   & 230  & 230  \\
$S/\sqrt B$ &  2.3 & 2.4 & 2.8 \\
\hline
\end{tabular}
\end{table}

\subsection{A 4 TeV muon Collider}

Achieving $WW$ subprocess energies above $1-2\tev$
is critical for studies of the SEWS physics, and is
only possible with high event rates at lepton-antilepton 
($\ee$ or $\mm$ colliders) or quark-antiquark (hadron collider)
subprocess energies of order $3-4\tev$. 

%
\begin{table*}[tbh]
\centering
\caption[]{\label{tableiv}\small
Total numbers of $W^+W^-, ZZ$ and $W^+W^+ \rightarrow4$-jet
signal ($S$) and background ($B$) events calculated for  a 4~TeV
$\protect \mm$ collider with  integrated luminosity 200~fb$^{-1}$
(1000~fb$^{-1}$ in the parentheses), for acceptance cuts as 
detailed in Ref.~19. 
%
%
The statistical significance $S/\sqrt B$ is given for the signal
from each model.
The hadronic branching fractions of the decays and the $W^\pm/Z$
identification/misidentification are included.}
\bigskip
\begin{tabular}{|l|c|c|c|} \hline
 & Scalar  & Vector & LET-K  \\
\noalign{\vskip-1ex}
& $m_H=1$ TeV  & $M_V=2$ TeV & ($m_H=\protect\infty$) \\
channels & $\Gamma_H=0.5$ TeV  & $\Gamma_V=0.2$ TeV & Unitarized\\ 
\hline\hline
$\mu^+ \mu^- \to \bar \nu \nu W^+ W^-$ &  &  & \\
$S$(signal) & 2400 (12000)  & 180 (890)  & 370 (1800)  \\
$B$(backgrounds)
& 1200 (6100)   & 25 (120)  & 1200 (6100)  \\
$S/\sqrt B$ & 68 (152) & 36 (81) & 11 (24) \\ \hline
$\mu^+ \mu^- \to \bar\nu \nu ZZ$& & &  \\
$S$(signal) &  1030 (5100)  & 360 (1800)  & 400  (2000)  \\
$B$(backgrounds)
& 160 (800)    & 160 (800)  & 160 (800) \\
$S/\sqrt B$ & 81 (180) & 28 (64) & 32 (71) \\ \hline \hline
$\mu^+ \mu^+ \to \bar\nu \bar\nu W^+ W^+$ & &  &  \\
$S$(signal) & 240 (1200) & 530 (2500) & 640 (3200)   \\
$B$(backgrounds)
& 1300 (6400)  & 1300 (6400)  & 1300 (6400) \\
$S/\sqrt B$ & 7 (15) & 15 (33) & 18 (40) \\ \hline
\end{tabular}
\end{table*}

\begin{figure}[tbp]
\centerline{\psfig{file=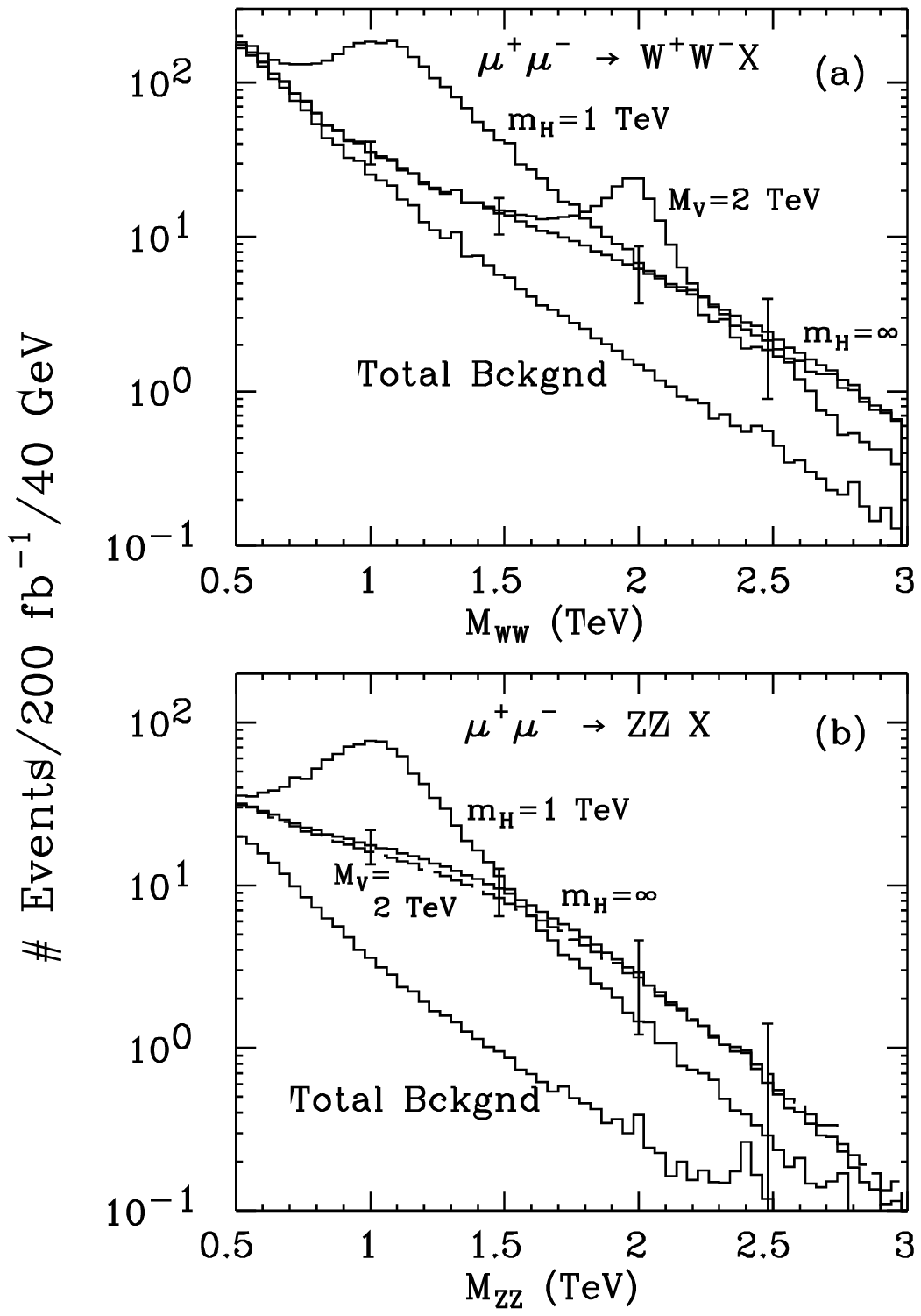,width=10cm}}
\begin{minipage}{12cm}       
\bigskip
\fcaption{{\small
Number of events at a 4-TeV muon collider
with $L=200\protect\fbi$ for SEWS models 
(including the combined backgrounds) and for the 
combined backgrounds alone in the 
(a) $\wp\wm$ and (b) $ZZ$ final states after imposing all 
acceptance cuts as detailed in Ref.~19. 
Sample signals shown are: (i) the SM Higgs with $m_H=1\tev$;
(ii) LET-K model (denoted by $m_H=\protect\infty$); 
and (iii) the Vector model with $M_V=2\tev$
and $\Gamma_V=0.2\tev$. Sample statistical uncertainties 
for the illustrated 40 GeV bins are shown
in the case of the LET-K model.}}
\label{mvvfullcuts}
\end{minipage}
\end{figure}

Consequently, a muon collider facility with center
of mass energy $\rts\sim 3-4\tev$ and luminosity
$L=200-1000\fbi$  allowing both $\mup\mum$ and $\mup\mup$
(or $\mum\mum$) collisions would be a remarkably powerful 
machine for probing the SEWS physics.  
Table~\ref{tableiv} presents the signal and background 
rates at a 4-TeV muon collider\cite{sewsmuon}. 
We see that
\begin{itemize}
\item the statistical significance of the SEWS signal
is high for all channels, regardless of model.
Even the $\wp\wm$ signal for LET-K and $\wp\wp$ signal
for $m_H=1\tev$ are clearly visible with only $L=200\fbi$, 
becoming thoroughly robust for $L=1000\fbi$;

\item
models of distinctly different types are easily distinguished
from one another, as previously discussed in the NLC study and
as more clearly demonstrated in Fig.~\ref{mvvfullcuts}(a)--(b).

\item
event rates for even the weakest of the model signals studied are such
that the $M(WW)$ distributions could be quantitatively
delineated, thereby providing a direct measurement of the underlying
strong $WW$ interaction amplitude as a function of the subprocess
energy and strong differentiation
among various possible models of the strongly interacting electroweak
sector;

\item
statistics are even sufficient that a model-independent
projection analysis can be applied to isolate the polarization
$TT$, $TL$ and $LL$ components of the cross section\cite{sewsmuon}. 
This would further delineate the correct theory
underlying the strong electroweak interactions based on 
the polarization studies for the final state $WW$.

\end{itemize}

\section{Summary}

A strongly interacting electroweak sector remains a logical possibility
responsible for the electroweak symmetry breaking. 
A phenomenological approach by adopting a (relatively) 
model-independent parameterization for such a sector is taken
to study the sensitivity to the SEWS physics at future colliders. 
We find that

\begin{enumerate}

\item at the LHC with  $\sqrt s=14\tev$, an observable
excess of events in the pure leptonic $W$ decay modes
can be seen for all of the strongly interacting models 
that we consider, after several years of running  with an annual
luminosity of  $100~{\rm fb}^{-1}$. The major problem for observing
SEWS effects is the large SM background and rather low pure leptonic
final state rates for the signals;

\item due to the clean experimental environment at an $e^+e^-$
collider, the hadronic $W$-decay modes may be used for searching
for SEWS $WW$ events. This leads to a large signal sample 
if the SEWS physics emerges near about 1 TeV region. On the
other hand, the available CM energy of about $1.5\tev$ may 
limit the SEWS physics reach, except for a vector dominant
model. The $e^-$ beam polarization could
help to enhance the signal observability. The option for
an  $e\gamma$ collider may be useful to uniquely produce other
states through $\gamma W\to A_1$ and $\gamma Z\to \omega_H^{}$;

\item
to fully explore $WW$ scattering physics at $1-2\tev$, 
a muon collider facility with center
of mass energy $\rts\sim 3-4\tev$ and luminosity
$L=200-1000\fbi$ would be a remarkably powerful machine.  
It not only yields sufficiently large event rates for SEWS
signals, but also allows detailed study for the signal structure
to uncover the possible underlying dynamics.

\end{enumerate}

\section{Acknowledgements}

I would like to thank Bernd Kniehl for his kind invitation and
hospitality extended to me during this stimulating workshop.
I thank J. Bagger, V. Barger, M. Berger, K. Cheung,
J. Gunion, Z. Huang, P.Q. Hung,
G. Ladinsky, R. Phillips, R. Rosenfeld and C.-P. Yuan
for collaboration which led to most of the results presented here.
I also thank the Fermilab Theory Group for the hospitality
when I finalize this report.
This work is supported in part by the U.S. 
Department of Energy Under contract DE-FG03-91ER40674, 
and by the Davis Institute for High Energy Physics.


\end{document}